\documentclass[12pt,preprint]{aastex}

\slugcomment{}

\shorttitle{Rotational Velocities of HVSs}
\shortauthors{L\'opez-Morales \& Bonanos}

\begin{document}

\title{``Slow'' and Fast Rotators Among Hypervelocity Stars\altaffilmark{1}}

\author{Mercedes L\'opez-Morales\altaffilmark{2,3} \& Alceste
Z. Bonanos\altaffilmark{2,4}}


\altaffiltext{1}{Based on data gathered with the 6.5 meter Clay Magellan
Telescope located at Las Campanas Observatory, Chile.}
\altaffiltext{2}{mercedes@dtm.ciw.edu, bonanos@dtm.ciw.edu. Carnegie Institution of Washington, Department of Terrestrial
Magnetism, 5241 Broad Branch Rd. NW, Washington D.C., 20015, USA}
\altaffiltext{3}{Hubble Fellow.}
\altaffiltext{4}{Vera Rubin Fellow.}

\begin{abstract}

We measure the projected rotational velocities of the late B-type
hypervelocity stars HVS7 and HVS8 from high resolution spectroscopy to
be 60 $\pm$ 17 km~s$^{-1}$ and 260 $\pm$ 70 km~s$^{-1}$. The 'slow'
rotation of HVS7 is in principle consistent with having originated in a
binary system, assuming a high inclination angle of the stellar rotation
axis. However, the fast rotation of HVS8 is more typical of single
B-type stars. HVS8 could have therefore been ejected by a mechanism
other than that proposed by Hills. We also estimate the effective
temperatures and surface gravities for HVS7 and HVS8 and obtain an
additional measurement of their radial velocities. We find evidence in
support of a blue horizontal branch nature for HVS7, and a main sequence
nature for HVS8.

\end{abstract}


\section{Introduction} \label{sec:intro}

The recent discovery of 10 hypervelocity stars \citep[HVSs;][]{Brown05,
Edelmann05, Hirsch05, Brown06a, Brown06b, Brown07} has raised many
questions about their nature and origin. The most widely accepted
ejection mechanism, proposed by \citet{Hills88}, involves the
encounter of a close binary with a supermassive black hole (SMBH). Other
possible mechanisms ejecting stars from the Galactic center involve
intermediate-mass black holes \citep[IMBHs; e.g.][]{Yu03, Lockmann08}, a
binary massive black hole \citep[BMBH; e.g.][]{Yu03, Merritt06,
Sesana06, Sesana07}, or a cluster of stellar mass black holes around the
SMBH \citep{Oleary08}. \citet{Hansen07} claimed that the rotational
velocities of HVSs should be lower than those measured for single stars
of the same spectral type if they originated in binaries, because of
tidal effects. He predicted that the rotational velocities of the known
B-type HVSs should be $\sim70-90$ km~s$^{-1}$, based on values compiled
by \citet{Abt04} for B-stars in binaries. \citet{Lockmann08} predicted
high rotational velocities for HVSs that were ejected by a very close
encounter with an IMBH in the Galactic center, however such encounters
are very unlikely.

These predictions cannot be tested with existing observations, as the
low resolution of the discovery spectra of most HVSs is not sufficient
to determine projected rotational velocities ($vsini$). The only HVS
with high resolution spectroscopy and a $vsini$ measurement is HE
0437--5439, found by \citet{Edelmann05}. It has $vsini=55 \pm
1$~km~s$^{-1}$ \citep{Bonanos08}, in agreement with the prediction of
\citet{Hansen07}. However, \citet{Bonanos08} and \citet{Przybilla08}
measured half-solar metallicity for this early B-star, establishing its
origin in the Large Magellanic Cloud (LMC). The possible ejection
mechanisms for this star include an interaction with an IMBH or a SMBH,
and a dynamical interaction of a single star in a dense cluster. This
example demonstrates the importance of high resolution spectroscopy
for understanding this newly discovered class of objects.

Of the remaining HVSs, HVS2 \citep[or US\,708;][]{Hirsch05}, is
classified as an evolved sdO star and reasonably well
understood. However there is some ambiguity in the nature of the late
B-type HVSs, since at their temperatures and gravities, the blue
horizontal branch (BHB) crosses the main sequence. Hot BHB stars
generally have low rotational velocities and peculiar chemical
abundances \citep{Behr03a}, thus high resolution
spectroscopy of these faint HVSs can determine their nature by
measuring their atmospheric parameters, chemical abundances and
$vsini$. In addition, time series photometry can reveal pulsations and
confirm their main sequence nature, as was done for HVS1 by
\citet{Fuentes06}.

Motivated by the lack of $vsini$ and stellar parameter measurements for
most of the known HVSs and the possibility of testing the nature
of the SMBH in the center of our Galaxy, we performed
high resolution spectroscopy of two HVSs. In this Letter we present our
results.

\section{Observations} \label{sec:obs}

We collected spectra of HVS7 and HVS8 (SDSS
J113312.12$+$010824.9 and J094214.04$+$200322.1) with the blue chip of
the MIKE spectrograph \citep{Bernstein03} installed at the 6.5-m
Magellan Clay Telescope at Las Campanas Observatory (Chile), on two half
nights on UT 2008 January 18--19. Each star was observed twice, with
individual exposure times between 900 and $1200\,s$, using a $1\arcsec
\times 5\arcsec$ slit and 3$\times$3 binning. The total exposure times
were $2100\,s$ for HVS7 and $2400\,s$ for HVS8. The resolution of the
spectra is R = 32,000 at 4500 \AA.

The spectra were extracted using the MIKE reduction pipeline
\citep{Kelson00, Kelson03}. The extracted spectra for each star were
then averaged, normalized and merged.  The wavelength coverage of the
 merged spectra is 3900-5050 \AA, with an average S/N-ratio per
pixel of 15 for HVS7 and 14 for HVS8, based on the extracted continuum
around 4500 \AA. These S/N-ratios and our spectral resolution are
sufficient to distinguish between high \citep[$>130-140$
km~s$^{-1}$;][]{Abt02} and low \citep[$70-90$
km~s$^{-1}$;][]{Hansen07} $vsini$ values for B-stars. Next, we
corrected the wavelength scale for Doppler shift, to allow comparison
of the spectra with models (see \S3). We measured the heliocentric
radial velocity of each star using the IRAF\footnote{IRAF is
distributed by the NOAO, which is operated by the Association of
Universities for Research in Astronomy, Inc., under cooperative
agreement with the NSF.}  cross-correlation package RVSAO
\citep{Kurtz98} and the grid of models described in \S
3. Table~\ref{tab:rv} lists our results and the values previously
reported by \citet{Brown07}.  \S4 discusses the implications of our new radial
velocity measurements.

\section{Spectral Analysis} \label{sec:spec}

Our high resolution spectra allow direct determination of the effective
temperature $T_{eff}$, surface gravity $logg$, and $vsini$ of the stars
by comparing synthetic model spectra to the observations. The S/N-ratio
of the data is however too low to reliably measure abundances.

We generated a grid of synthetic spectra using the LTE ATLAS9 models and
opacities developed by \citet{Kurucz93}. The grid covers $T_{eff}$
between 8000--15000~K in steps of 1000~K, and $logg$ between 3.0--5.0 in
steps of 0.25 dex. The metallicity was set to solar,
assuming that the HVSs are ejected from the Galactic center, where
abundances are solar or supersolar \citep{Carr00, Ramirez00, Najarro04,
Wang06, Cunha07}.  For the macro- and micro-turbulence velocities we
adopted 0 and 2 km~s$^{-1}$, which are typical for late
B-stars \citep{Fitzpatrick99}. The models were broadened by 0.15 \AA~ to
match MIKE's instrumental profile and resampled to a dispersion of 0.03
\AA/pix to match the dispersion of the stellar spectra. Finally, we
convolved each model with rotational profiles between 10--350
km~s$^{-1}$ in 10 km~s$^{-1}$ velocity increments.

Simultaneous fits to $T_{eff}$, $log g$ and $vsini$ were performed for
each star by iteratively comparing each model to the data. The
agreement between each model and the observed spectra is quantified by
the spectroscopic quality-of-fit parameter, $z$ (normalized
$\chi^{2}$), defined by \citet{Behr03a} and given by the equation
\begin{equation}
 z=\sqrt\frac{N_{points}}{2} \left( \frac{rms^{2}}{rms^{2}_{min}} - 1\right),
\end{equation}

\noindent where $N_{points}$ is the number of points in the spectrum,
and $rms$ and $rms_{min}$ are the root mean squared 
deviation between each model and the stellar spectrum, and the smallest
value of the rms found. $z$ = 0 gives the best model fit, and $z$ = 1
defines the statistical 1$\sigma$ confidence interval of the result. The
following subsections describe the derivation of $T_{eff}$, $log g$ and
$vsini$ for each target.

\subsection{HVS7} \label{sec:hvs7}

The spectrum of HVS7 (V=17.80 mag, S/N=15) includes four Balmer lines
($H_{\beta}$ -- $H_{\epsilon}$) from which $T_{eff}$, $logg$ and
$vsini$ can be estimated. We also detect \ion{Mg}{2}, \ion{Si}{2},
\ion{Ca}{2} and \ion{Fe}{2} lines that can in principle be used to
further constrain $T_{eff}$ and $vsini$, however the two
main $T_{eff}$ indicators (\ion{Mg}{2} and \ion{Si}{2}) have anomalous
line strengths and cannot be used to constrain the $T_{eff}$. We are
therefore left with only the Balmer lines that are simultaneously
sensitive to $T_{eff}$, $logg$ and $vsini$, but can still provide
non-degenerate values of these parameters for late B-type stars
\citep[see \S 4 of][]{McSwain07}. We performed two tests to verify
that Balmer lines alone are sufficient to simultaneously derive the
three parameters: a) we applied our analysis to
synthetic spectra with added random noise matching the S/N-ratios of
the observations, and in all cases recovered the input values within
errors, b) we applied our analysis to a high S/N spectrum of a late
B-type star (HR7447, B5 III), kindly provided by L. Lyubimkov. Our
analysis yielded $T_{eff}$ = 14,000 $\pm$ 1000 K, $logg$ = 3.75 $\pm$
0.25 dex, and $vsini$ = 70 $\pm$ 20 km~s$^{-1}$, in agreement with the
parameters derived by \citet[][$T_{eff}$ = 13,400 K, $logg$ = 3.64 dex,
$vsini$ = 76 km~s$^{-1}$]{Lyubimkov02, Lyubimkov04}.

\begin{figure}[hbt]
\epsscale{1.0}
\includegraphics[angle=90, width=6.5in]{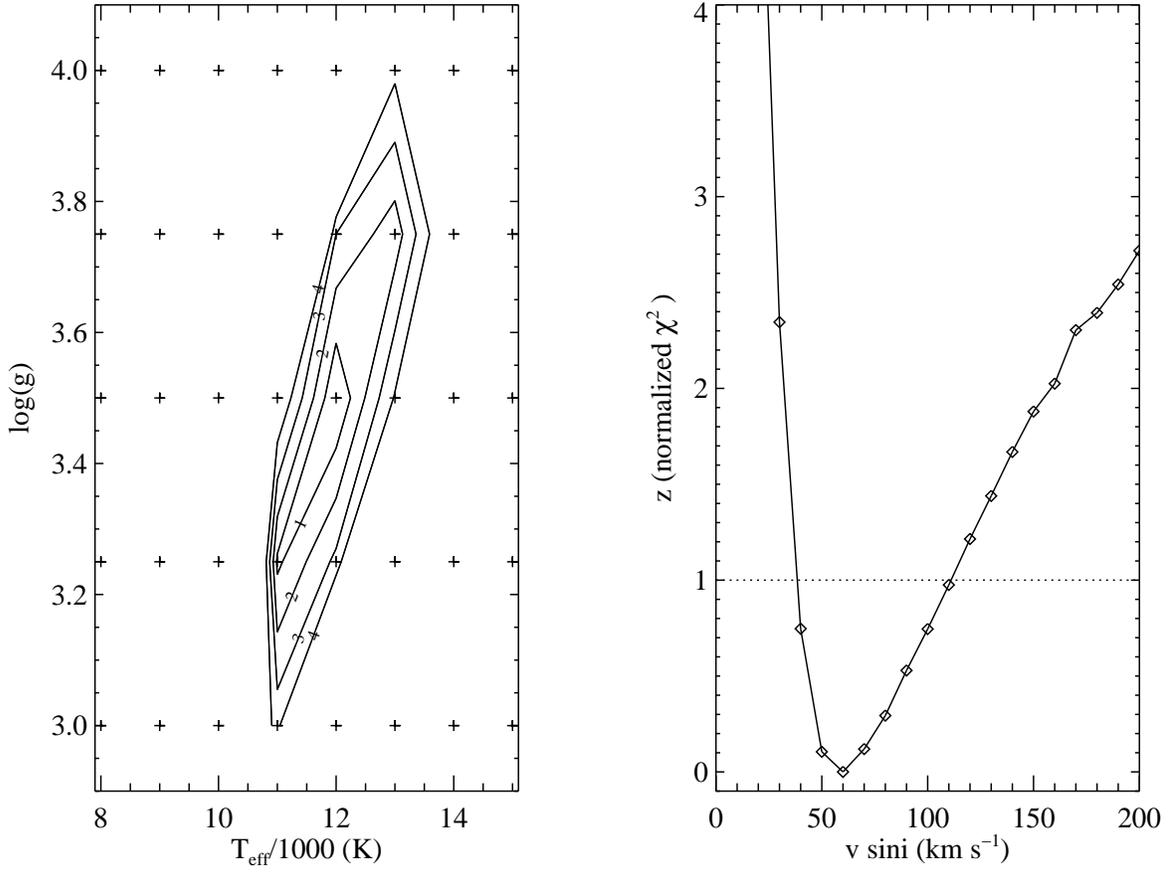}
\caption{$z$ minimization result for full spectrum of HVS7. {\it
Left:} Contour plot of $logg$ vs. $T_{eff}$ for a fixed $vsini$ of 60
km~s$^{-1}$; contours correspond to $z=1,2,3,4$; crosses to the models
in the Kurucz grid. {\it Right:} $z$ vs. $vsini$ for the best
fit values $T_{eff}$ = 12,000 K and $logg$ = 3.50 dex. $z$ = 1 shows the statistical 1$\sigma$ error of the fit;
the result is $vsini=60\pm17$ km~s$^{-1}$. A similar analysis was done for HVS8.}
\label{fig:conts}
\end{figure}

\begin{figure}[hbt]
\epsscale{1.0}
\plotone{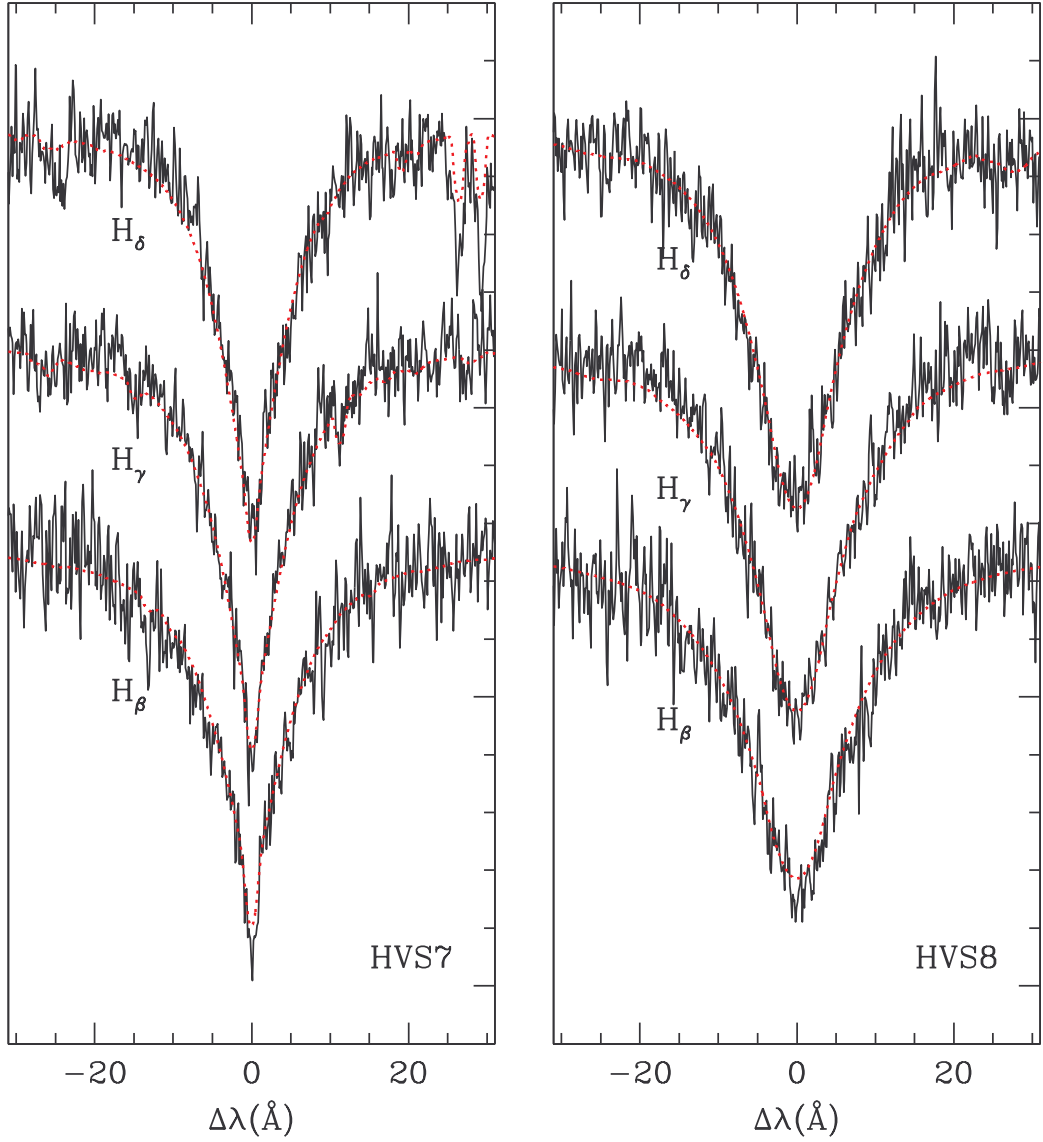}
\caption{Comparison of the best fit models to the $H_{\beta}$,
$H_{\gamma}$ and $H_{\delta}$ Balmer lines of HVS7 $(left)$ and HVS8
$(right)$. Models are shown as red (dotted) lines.}
\label{fig:Hlines}
\end{figure}

We then proceeded to fit the spectrum of HVS7 for $T_{eff}$, $logg$
and $vsini$. We performed several tests to determine the stability of
the best fit solutions.  We ran fits to the entire spectrum, 100~\AA~
windows centered on each Balmer line (to ensure that the wings and
some continuum are included), portions of the spectrum outside the
Balmer lines, and 10--20~\AA~ windows centered on metal lines.  In the
last two cases we had to fix $T_{eff}$ and $logg$ to the values from
the fits to the entire spectrum and the Balmer lines and only fit for
$vsini$. All the tests give fully consistent results, with the
following best fit parameters: $T_{eff}$ = 12,000 $\pm$ 1000 K, $logg$
= 3.50 $\pm$ 0.25 dex, and $vsini$ = 60 $\pm$ 17 km~s$^{-1}$. The $z$
minimization results for the full spectrum are shown in Figure
\ref{fig:conts}. We have adopted conservative errors for $T_{eff}$ and
$logg$ equal to the grid step size, versus their smaller 1$\sigma$
statistical errors. The statistical 1$\sigma$ errors for $vsini$
(horizontal dotted line in the $z$ vs. $vsini$ plot in Figure
\ref{fig:conts}), are $-20$ and $+50$ km~s$^{-1}$, however visual
comparison of the models to the observed spectrum show they are too
large. Instead we adopted the errors resulting from the fits to
individual metal lines.  The left panel in Figure \ref{fig:Hlines}
compares the best fit model to the Balmer lines of HVS7.

\begin{figure}[hbt]
\epsscale{1.0}
\plotone{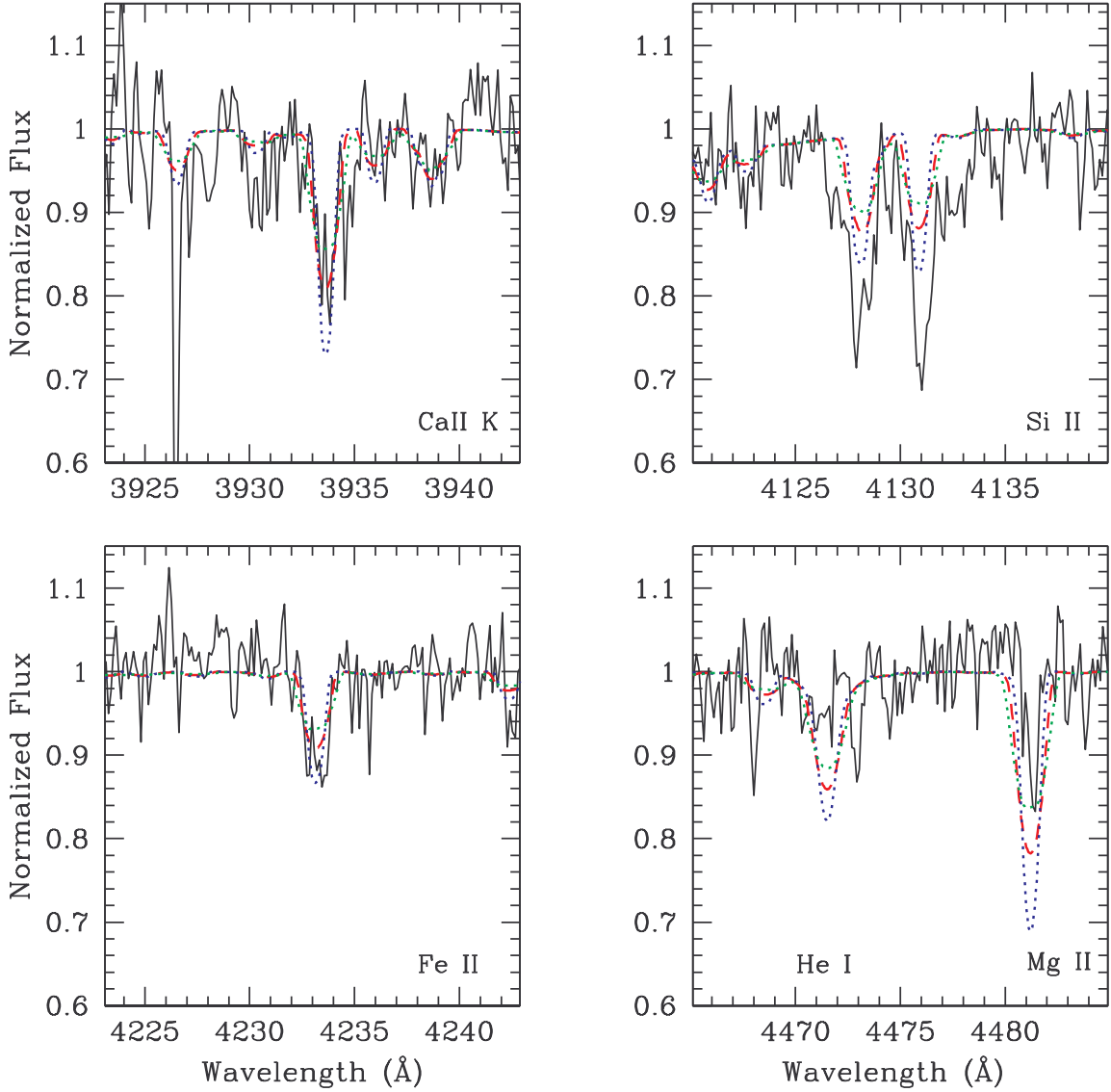}
\caption{Metal lines detected in HVS7; the overplotted models
correspond to $T_{eff}$ = 12,000 K, $logg$ = 3.50 dex and $vsini$ = 40
km~s$^{-1}$ (blue), 60 km~s$^{-1}$ (red), and 80 km~s$^{-1}$
(green). \ion{Ca}{2} $\lambda$3933 and \ion{Fe}{2} $\lambda$4233 are
well reproduced by our best fit model (red), while \ion{Si}{2}
$\lambda\lambda$4128-30, \ion{He}{1} $\lambda$4471 and \ion{Mg}{2}
$\lambda$4481 show abundance peculiarities. The sharp absorption line
at 3926.5 \AA~ corresponds to interstellar \ion{Ca}{2} K after
correcting the radial velocity of the HVS7 spectrum to the
heliocentric reference system.}
\label{fig:metals}
\end{figure}

Figure \ref{fig:metals} shows metal lines detected in HVS7 with
$vsini$ = 40, 60, and 80 km~s$^{-1}$ models overplotted.  The purpose
of this plot is two-fold; the left-side panels show how the
\ion{Ca}{2} K and \ion{Fe}{2} $\lambda$4233 lines ($vsini$ is derived
from these two and the \ion{Fe}{2} $\lambda\lambda$4549 and 4583
lines) give $vsini$ = 60 km~s$^{-1}$ as the best model fit, and their
depths agree with the solar abundance adopted in the models. The
$vsini$ from these lines also agrees with the fits to the full
spectrum and the Balmer lines. The right-side panels show the behavior
of the \ion{Si}{2} doublet, the \ion{Mg}{2} and \ion{He}{1}
lines. None of the models in the grid can reproduce the depths of
those lines. While \ion{He}{1} and \ion{Mg}{2} seem depleted in the
atmosphere of HVS7, the \ion{Si}{2} 4128/4130 lines seem strongly
enhanced. The models cannot reproduce either the depth nor the line
strength ratio of the \ion{He}{1} and \ion{Mg}{2} lines. The
enhancement of the \ion{Si}{2} 4128/4130 doublet is even more
significant when taking into account that the Kurucz LTE ATLAS9 models
overpredict the strengths of these lines. This problem persists even
after including non-LTE corrections \citep{Smartt01} . Abundance
peculiarities have been noticed before in BHB stars by several authors
\citep{Glaspey89, Moehler99, Behr99}; however, most of these are very
slow rotators \citep[$<8$ km~s$^{-1}$;][]{Behr03a}.

\subsection{HVS8} \label{sec:hvs8}

The spectrum of HVS8 (V=18.09 mag, S/N = 14) has a S/N similar to
the HVS7 spectrum, however, inspection of the spectrum of HVS8 for metal lines
gives null results. This can be explained by very low metal abundances,
strong depletion, or highly broadened metal lines.  The $vsini$ obtained
below points towards the latter case.

To derive $T_{eff}$, $logg$ and $vsini$ we used only the spectrum
above 4000 \AA~ because of problems with the continuum normalization
at shorter wavelengths. As with HVS7, we simultaneously fit for
$T_{eff}$, $logg$ and $vsini$ by iteratively comparing the spectrum of
HVS8 to our model grid. We ran fits to the entire spectrum and
160~\AA~ windows centered on the Balmer lines, which gave consistent
parameters: $T_{eff}$ = 11,000 $\pm$ 1000 K, $logg$ = 3.75 $\pm$ 0.25
dex, and $vsini$ = 260 $\pm$ 70 km~s$^{-1}$. The lack of metal lines
in the spectrum of HVS8 is consistent with a high $vsini$ that results
in strong line broadening. The error in $vsini$ in this case comes
directly from the $z$ = 1 statistical 1$\sigma$ result, as visual
comparison of the spectrum to the models does not allow us to place a
finer constraint. The right-side panel in Figure \ref{fig:Hlines}
compares the best fit model to the $H_{\beta}$, $H_{\gamma}$ and
$H_{\delta}$ lines of HVS8. The flat-bottomed cores of the Balmer
lines, which are the most sensitive regions to $vsini$, clearly
illustrate that HVS8 rotates faster than HVS7.

\section{Radial Velocities}\label{sec:rv}

The new radial velocity observations in Table \ref{tab:rv} provide a
third epoch for each star and allow to check for variations. Our
radial velocity measurement for HVS7 is identical to the values
reported by \citet{Brown06b} and \citet{Brown07}, within errors. Such
measurements provide clues to the nature of HVSs. As pointed out by
Brown et al., determining the nature of late B-type HVSs is not
straightforward because late-type main sequence B-stars and hot Blue
Horizontal Branch (BHB) stars have identical atmospheric
parameters. BHB stars are less luminous and therefore closer in
distance to us. Establishing the evolutionary stage of
HVS7 is critical because its radial velocity is marginally consistent
with it being a BHB runaway star bound to our Galaxy
\citep{Brown06b}. The lack of significant radial velocity variations
for HVS7 suggests it is not a binary, nor a pulsator. Slowly pulsating
main sequence B-type stars typically show radial velocity variations
of $\sim$ 20~km~s$^{-1}$ in amplitude \citep{Aerts99,Mathias01}, while
BHB stars appear to be stable, as they fall outside the RR Lyrae
instability strip \citep{Contreras05, Catelan06}. The long term radial
velocity stability of HVS7, combined with the metal abundance
anomalies (see \S\ref{sec:hvs7}), hint towards HVS7 being a BHB
star. Its $vsini$ (60$\pm17$ km~s$^{-1}$) is higher than typically
found for BHB stars ($<8$ km~s$^{-1}$), although rotators with $vsini$
up to $\sim40$ km~s$^{-1}$ have been observed \citep[e.g.][]{Behr03a,
Behr03b}. The true nature of HVS7 as a bound BHB star will have to be
disentangled by astrometry.

For HVS8 we detect a radial velocity variation of 23 km~s$^{-1}$,
consistent with a pulsating main sequence B-type star.  We cannot
discard the possibility of HVS8 being a binary, although the system
would have a very low mass-ratio, since there is no evidence of lines
from a companion in the spectrum. The star is most likely a main
sequence slow pulsator, like HVS1 \citep{Fuentes06}. The low S/N of
our spectrum does not allow to test for metal abundance anomalies
in HVS8, however, the high rotational velocity of this star will make
its abundance analysis difficult, even with higher S/N spectra.

\section{Discussion} \label{sec:sum}

We have derived $vsini$, $T_{eff}$ and $logg$ for HVS7 and HVS8, two
of the ten currenty known HVSs. Their $T_{eff}$ and $logg$ are
consistent with the stars being late B-type, as initially classified
by \citet{Brown06b, Brown07} using photometric color indexes. HVS7 has
a projected rotational velocity $vsini$ = $60\pm17$ km~s$^{-1}$, while
for HVS8 $vsini$ = $260\pm70$ km~s$^{-1}$.  These measurements provide
the first direct observational test to the prediction by
\citet{Hansen07}, who suggests that HVSs ejected via Hills' mechanism
should rotate systematically slower ($70-90$ km~s$^{-1}$) than single
stars of the same spectral type \citep[$134\pm7$
km~s$^{-1}$;][]{Abt02}. If the HVSs have fast rotational velocities
typical of single B-type stars in the field, other ejection
mechanisms, such as three-body encounters of single stars with $\sim$
$10^{3}$--$10^{4}$ $M_{\sun}$ IMBHs, with a binary MBH, or with $\sim$
10 $M_{\sun}$ stellar-mass black holes orbiting the Galactic SMBH,
have to be invoked.

The $vsini$ values of HVS7 and HVS8 are lower limits to their true rotational
velocities, imposed by the inclination angle of the rotation axis of the
stars. If the inclination of the rotation axis of HVS7 is low, its
rotational velocity could in principle be much higher. In that case both
targets are inconsistent with Hansen's prediction for Hills'
scenario. However, a sample of only two $vsini$ measurements is not
enough to conclusively discern between the different scenarios proposed
and more $vsini$ measurements are necessary.  Statistical tests
performed by \citet{Perets07} conclude that a sample of 25 or more HVSs
will be needed to distinguish between scenarios at a $\ge$ 95$\%$
confidence level.

We also detect abnormal enhancement and depletion effects in
the strength of some of the metal lines of HVS7. These anomalies,
together with the apparent lack of pulsations are consistent with HVS7
being a BHB star. Its $vsini$ is also marginally consistent with it
being a fast-rotating BHB star, if the inclination angle of the star's
rotation axis is close to 90 degrees. However, confirmation that HVS7 is
a BHB star will not be possible until its proper motion is accurately
measured. Finally, we find evidence of radial velocity variations in
HVS8 consistent with a pulsating main-sequence B-type star
nature. Additional radial velocity measurements and time-series
precision photometry will confirm this detection.

\acknowledgments{We thank N. Morrell for doing the
observations, I. Ribas for his script to merge echelle
orders, and W. Brown for suggestions.  MLM acknowledges support
provided by NASA through Hubble Fellowship grant HF-01210.01-A awarded
by the STScI, which is operated by the AURA, Inc. for NASA, under
contract NAS5-26555. AZB acknowledges support from
the Carnegie Institution of Washington through a Vera Rubin Fellowship.}



\clearpage

\begin{deluxetable}{cccc}
\tablewidth{0pc}
\tablecaption{\sc Heliocentric Radial Velocities}
\tablehead{
\colhead{Target} & \colhead{RV (km~s$^{-1}$)} & \colhead{RV (km~s$^{-1}$)}\\
\colhead{} & \colhead{(this work)} & \colhead{(previous work\tablenotemark{*})}} 
\startdata
HVS7 & 529 $\pm$ 2 & 531 $\pm$ 12  \\
HVS8 & 489 $\pm$ 2 & 512 $\pm$ 10  \\
\enddata
\label{tab:rv} 
\tablenotetext{*}{\citet{Brown07} report two radial velocity
observations for each star and find identical values within their
errorbars.}
\end{deluxetable}


\end{document}